\documentstyle{l-aa}
\begin{document}
\thesaurus{08.01.1, 08.16.3, 10.07.3 47 Tuc, 10.08.1, 12.03.3}
\title{Lithium observations in  47 Tuc
\thanks{Based on observations collected at ESO, La Silla}}
\author{L. Pasquini \inst{1} 
\and P. Molaro \inst{2} }
\offprints{Luca Pasquini}
\institute{ European Southern Observatory, Casilla 19001, Santiago 19, Chile
\and
Osservatorio Astronomico di Trieste, Via G.B. Tiepolo 11 I34131,
Trieste, Italy.}
\date{Received; Accepted}
\maketitle
\begin{abstract}
We present high resolution  observations (FWHM=15 km s$^{-1}$)
of the Li
region in 5 stars belonging to the old ($\approx$14 Gyrs), metal rich 
([Fe/H]$\approx$-0.7) globular cluster 47Tuc.
At the ESO NTT telescope 
we  obtained EMMI spectra for three V$\sim$17.4
 magnitude stars  located at the turnoff with T$_{eff}$ on the 
Spite {\it plateau}. 
In two of the turnoff stars the lithium line is
clearly detected and the  mean lithium 
content [Li] = 2.37$\pm$ 0.08$\pm$0.07 is derived. 
 This value  is 
slightly higher than
 the Spite {\it plateau} at [Li]=2.19$\pm 0.016$ for field 
halo stars obtained with the same T$_{eff}$ scale by Bonifacio
and Molaro (1996).

For the third turnoff star 
 only an upper limit could be derived: [Li]$<$ 2.19. This upper limit may 
 suggest the presence of some dispersion on the plateau stars of 47Tuc 
as it is observed in field stars of similar metallicity.

When compared with field dwarfs  of [Fe/H] $\approx$ -0.7  the Li content
of 47Tuc turnoff stars  is among the highest observed. Considering
the  extreme age of 47Tuc this suggests that  no significant depletion 
affected  these stars and it  implies 
  a mild Li Galactic enrichment  between  the epoch of Pop II formation 
and the formation of 47Tuc stars with metallicities
up to  $\sim -0.7$. Spallation reactions, as deduced from Be observations,
can account for such an increase.
   
The remaining two stars are evolved, and only upper limits could be 
placed at [Li] $<$0.7 and $<$-0.7. These very low values imply some
extra depletion mechanism in addition to dilution as it is 
observed in evolved field stars by Pilachowki et al (1993) 
and in NGC6397 by Pasquini and Molaro (1996). 


\end{abstract}

\section{Introduction}

 The study of light elements, Li and Be and B, 
has progressed impressively
in the last years, both in the theoretical and observational aspects
(see e.g. the proceedings edited by Crane 1995). 

On the observational side, observations of Li in  large, well 
defined samples, are becoming  available,
(Brown et al. 1989, Pasquini et al. 1994, 
Randich et al. 1996), 
as well as new detailed observations 
in open clusters of different ages, either young (Soderblom et al. 1993)
or as old as the Sun (Balachandran 1995).  

For the study of the oldest population of the Galaxy, a large effort 
has been performed to analyze very metal poor stars in the field
(see i.e. Spite 1995 for a review), and 
the first data on Globular Clusters have become available, 
for the  old, metal poor GC's NGC6397 
(Molaro and Pasquini 1994, Pasquini 
and Molaro 1996) and M92 (Deliyannis et al. 1995).

On the theoretical side several  models have been developed which include
new important features.
Standard models have been built 
with a more realistic treatment of the mixing in the 
convection zone (D'Antona and Mazzitelli 1994), 
or  with  improved opacities 
(Swenson 1995). 
Non standard models  were also improved and they include 
mechanisms previously neglected or only partially treated, like 
rotational mixing (Pinsonneault et al. 1992, Zahn 1994), meridional 
circulation (Michaud and Charbonneau 1991), diffusion and
mass losses (Swenson 1995, Vauclair and Charbonnel 1995).

The homogeneity of stars in G.C. and the knowledge
of their overall characteristics offer a powerful tool 
to address several  topics concerning Li abundance. 
By studying, for instance, the dependence  of Li abundance on stellar
 mass in a given GC, a detailed comparison with theoretical models can 
be performed, allowing to disentangle between the different mechanisms 
operating in the
stellar interiors. By studying the Li abundance in clusters spanning a 
large range of ages
it would be possible to obtain precious information about the Galactic
evolution of lithium (Hobbs and Pilachowski 1988) and its primordial 
fraction (Trimble 1991).

One interesting target among the near GC is certainly 47Tuc 
(=NGC104 = C0021-723).
The cluster is  well studied; with an age of at least 13$\pm$ 0.5  Gyr
it belongs to the {\it old}  generation of G.C. although its 
metallicity is more 
than 30 times higher than other coeval clusters 
(Hesser et al. 1987). 
Observations of Li in 47Tuc and their comparison with the observations
in NGC 6397 and M92 are therefore particularly suitable for the
study of the Lithium behavior with respect to stellar 
metallicity at a given age.

\section{The observations}

The observations were obtained
between October 11 and October 14, 1994
using the EMMI spectrograph (Giraud 1995) at 
the Nasmyth focus of the 3.5m  NTT telescope. 
EMMI  was 
used in echelle mode, with grism 6 as crossdisperser
providing a spectral coverage of 2000 {\AA}.
With a Tectronix 2040$\times$2048 CCD and the  F/5 Long Camera
the scale  is  0.067 {\AA}/pixel in the
Li region. 
Due to variable metereological conditions and to 
 rather poor seeing, a slit aperture of 1.5
 arcsec was mostly used, which gives a resolving power of $\sim$ 18000. 
The slit height, fixed at  10 arcsec, gives enough (35) pixels
 perpendicular to the dispersion  direction to allow a good
background-sky subtraction. 
The CCD was binned by two along the  
resolution, in order to minimize
the contribution of the CCD Read Out Noise. The CCD was also read in  
`superslow' mode which gave a RON of about 4e$^-$/pixel. 

We were able to observe a total of 5 stars in 47Tuc,  from 
V=12.2 down to V=17.35. 

 Observations are summarized in Table 1. 
 The data were reduced 
using MIDAS (Banse et al. 1988) facilities. 
Special care was taken in the reduction of these low S/N data, in order 
to precisely subtract  the background (interorder+sky) contribution
and to take into account the presence of high energy particles events. 
To check the accuracy of the data the reduction was carried
 out using different procedures. 

The final spectra are presented in Figure 1. After rebinning to equal 
(0.12 {\AA}) wavelength step, 
the spectra were  coadded with the proper weight; note that no major 
spectrograph shifts were recorded during the whole run.   

The continuum was traced 
measuring the intensity of the spectra in several windows  
 in selected  regions surrounding the Li line.
In the case of very metal poor stars, this procedure can be easily applied
because of the featureless appearence of the spectra. At the metallicity 
of 47Tuc ([Fe/H] = -0.7), however, the selection of the windows must be done 
carefully, due to the presence  of metallic lines in the spectrum. 
The continuum regions were therefore selected from the inspection 
of  high resolution, high S/N spectra (Pasquini et al. 1994). 
  
Equivalent widths were 
measured with simple integration. The FWHM of the Li and Ca I (6717) lines
was of $\sim$ 0.34 {\AA},  in perfect agreement with what was expected 
from the instrument resolution.  
In estimating the errors on  measured equivalent widths, we have assumed 
a conservative approach, in view of the uncertainties 
which influence the accuracy of the data: $\Delta$EW = RMS*RES*W$^{1/2}$,
where RMS is the inverse S/N ratio  as measured in the continuum windows 
close to the Li line, RES the measured resolution and W the 
number of resolution elements on which the integration has been carried out.  
To these uncertainties a value up to $\sim$ 6 m{\AA} should be added due to 
the possible uncertainties in positioning the continuuum (Cayrel 1988). 

Equivalent widths and S/N ratio are given in Table 3. 

\begin{figure}
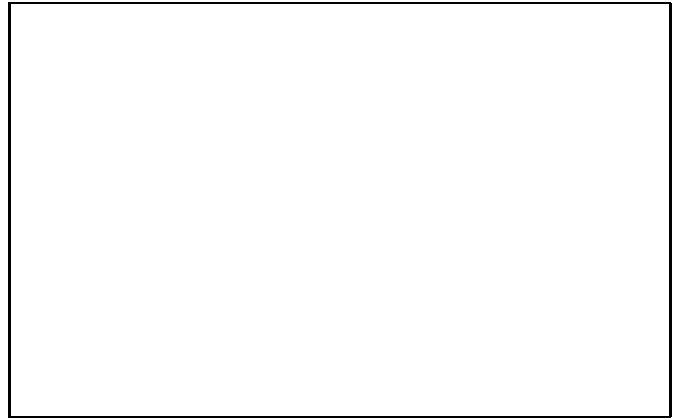

\picplace{5.5cm}
%\vbox{\psfig{figure=paper2fig1.ps,width=0.cm,
%            bbllx=2.0cm,bblly=2.0cm,bburx=4.0cm,bbury=4.0cm,clip=}}\par
%\psfig{figure=paper2fig1.ps,width=8.0cm,bbllx=65pt,bblly=90pt,bburx=800pt,bbury=680pt,clip=}
%\psfig{figure=prepfig1.ps,width=12.0cm,silent=}
\caption[1]{EMMI spectra of the 5 observed stars  around  the Li 6708 
\AA~~ region; the spectra were artificially displaced for
better viewing. The first spectrum is the composition of the two turnoff stars
(b5 and b7), where Li was clearly detected.}
\end{figure}

\section{Stellar parameters}

At a distance of $\sim$ 4.5 Kpc,
47Tuc is one of the best studied  globular clusters. It is located at high Galactic latitude and 
the reddening is rather low.  In the literature reddening estimates
range from E(B-V)=0.00 of Cannon (1974)  up to values of 0.08 
(Menzies 1973). However,  best evidence is for intermediate values: 
Crawford and Snowden (1975) derived E(B-V)=0.029$\pm$0.004;  
Hesser and Philip (1976) 
obtained E(B-V)=0.024$\pm$0.021 and Lee (1977) 
obtained E(B-V)=0.04$\pm$0.01. In the following we will adopt
 E(B-V)= 0.04 as suggested by Hesser et al. (1987), but an error of 0.02
is  possible here, and this is rather important in deriving the absolute value for Li in the cluster.
After a controversial discrepancy between the spectroscopic and 
photometric results, the last estimates of metallicity 
converge between the [Fe/H]=$-$0.65, derived photometrically 
by Hesser et al. (1987) and the [Fe/H]=$-$0.8 derived 
spectroscopically   by D' Odorico et al. (1985).
In the following, a value of [Fe/H]= -0.7  will be adopted, but note that 
the uncertainties in  metallicity have no significative influence in the
determination of the  Li abundances. 

The observed stars are listed in Table 1.  Finding charts were provided by 
Briley et al. (1994), together with 
stellar colours and magnitudes, which  are summarized  
in Table 2.
With a visual magnitude around V$\sim$ 17.35 and rather blue colors, 3  stars 
lie at the bright blue edge of the  turnoff, indicating that they are in  
the upper part of the main sequence. 

To determine the stellar effective temperature,  
we have adopted the temperature scale of Alonso et al. (1996b), based on the 
IRFM (see Alonso et al. 1996a), to be able to compare the
47Tuc abundances directly with the most recent determination of the Li 
{\it plateau} (Bonifacio and Molaro 1996) which is based on the 
Alonso et al. (1996a) effective temperatures.  
We note  that the effective temperature used are in a very 
good agreement (within 50 K) with those obtained by using the Briley 
et al. (1994) scale and with the Vandenberg and Bell (1985) scale used for 
NGC6397 by Pasquini and Molaro (1996). 

For star L2734, the effective temperature was derived from Buser and 
Kurucz (1992) ,
because none of the over mentioned scales is suitable for such a cool,
evolved object.
The uncertainty of this determination is somewhat higher than the 
one for the turnoff stars, but we note that the effective temperature 
adopted is not critical for this cool star.

\begin{table}
\caption{Summary of the observations: in column 1 the star names are 
given, the b suffix corresponds to stars from Briley et al. 1994, the L 
suffix for stars from Lee  (1977))}
\parskip0.2cm
\begin{center}
\tabcolsep0.2cm
\begin{tabular}{r|r|r}\hline
\multicolumn{1}{c}{Name} &
\multicolumn{1}{c}{Int. Time} &
\multicolumn{1}{c}{Comments}\\ \hline\hline
b7      & 5$\times$2 H   & R=18000 Poor weather Conditions \\
b6      & 3$\times$1.5 H & R=23000 \\
b5      & 2$\times$2 H  &  R=18000 \\
b1      & 2$\times$1.5 H&  R=18000 \\
L2734   & 2$\times$0.5 H&R=18000 \\
\hline
\end{tabular}
\end{center}
\end{table}

\begin{table}
\caption{Photometric data for the observed stars, `b' indicates 
stars from Briley et al. (1994), `L' from Lee (1977).}
\parskip0.2cm
\begin{center}
\tabcolsep0.2cm
\begin{tabular}{r|r|r|r|r}\hline
\multicolumn{1}{c}{Star} &
\multicolumn{1}{c}{V} &
\multicolumn{1}{c}{(B-V)} &
\multicolumn{1}{c}{(B-V)$_o$} &
\multicolumn{1}{c}{V$_r$ (Km/sec)}
\\ \hline\hline
b7       & 17.38  & 0.57 & 0.53 &  -16  \\
b6       & 17.35  & 0.59 & 0.55 &  -24  \\
b5       & 17.35  & 0.59 & 0.55 &  -17  \\
b1       & 16.18  & 0.83 & 0.79 &  -34  \\
L2734    & 12.21  & 1.43 & 1.39 &  -16  \\
\hline
\end{tabular}
\end{center}
\end{table}

From our spectra it was also possible to derive radial velocities for the 
observed stars. They were computed mostly by using the Ca I 6717 line, 
but also checked using other lines in the spectra. The results are  given 
in Table 2. No radial velocity 
standards were observed during our run, therefore the absolute values of
the measured V$_r$ can have a systematic zero  offset by a value 
probably up to 5 Km/sec. Due to
the possibility of checking the wavelengths of sky lines, instead, 
our internal accuracy is better than 2 Km/sec. As  can be seen, 
two of the stars (b1 and b6) differ by a significant amount (7 and 17
Km/sec respectively) from  the others, which are in good agreement
with the {\it mean} systemic velocity of the cluster of -19.4 Km/sec 
(Mayor et al. 1984). 
The observed  stars belong to a field  centered at a distance of
$\sim$ 11 arcmin from the core of the cluster, almost perfectly 
alligned to the west (Briley et al. 1994).
Mayor et al. (1984) showed that, due to the cluster dynamics,
in 47Tuc the mean radial velocities and  
the velocity dispersion vary with the distance from the 
cluster center and with the orientation. The measured 
dispersion at 11 arcmin from the core is about 7 Km/sec, and the expected 
velocity towards the west is very close to the the mean cluster value. 
The  velocity measured for star b6 is therefore 
well compatible with the Mayor et al. (1984) results, while for b1 
the membership is 
only marginal (2$\sigma$). Some additional spread among the observed stars
may be expected,  considering that their 
distance  from the cluster center differs by 7 arcmin.

Considering that  the  position of b1 in the color 
magnitude diagram fits  well the cluster mean curve,
we retain at this stage also b1  as a possible cluster member. 
Note also that the membership of b1 is not at this stage very relevant
for the following discussion, which is mainly centered on the turnoff stars.

\section{Li abundances}

The Lithium line is clearly present in two of the 3 turnoff  stars observed. 
In the spectrum of the  turnoff star b6,  no clear 
feature is present at the lithium rest wavelength, but a possible absorption 
line of $\sim$40 m{\AA} is present 0.1 {\AA} blueward from the nominal 
wavelength. We set this value as an upper limit to the Li equivalent width.  

Abundances are computed by using synthetic profiles generated by the SYNTHE code and atmospheric models  from the Atlas 9 ( Kurucz 1993). Convection is 
treated with the mixing length theory, with a scale height over 
pressure scale of 1.25, but without 
the overshooting option. The resulting COG's are very similar to those 
obtained with the old Kurucz and Bell and Gustafsson  model
(see the discussion in Molaro et al. 1995). 
If the Kurucz (1993) grid with overshooting would be used, the 
Lithium abundances presented here will increase by $\approx$ 0.05
   dex. Microturbulence was fixed to 1.0 Km/sec.

The Fe 6707.43 {\AA} line is not resolved in our spectra,
  but its contribution  was considered in the  computation. At the 47Tuc
metallicity and for the range of temperatures analyzed the FeI line has an
equivalent width  of  about 1-2 m{\AA}.
 
Because of the similarity among the turnoff stars, and the relatively 
low S/N ratio of the observations, we have decided to coadd the spectra
of the two turnoff stars, b5 and b7, with a clear detection of Li. 
The coadded spectrum is also  shown in Figure 1 and it has a mean S/N ratio 
of 50. In this spectrum the Li line has an equivalent width of 
52 m{\AA} and the Ca I line of 71 m{\AA}. For a 
5866 K star these values correspond to  [Li]=2.37   and [Ca/H]=-0.33
and -0.54 , with a microturbulence of 1 and 2 Km/sec respectively.  This is
consistent
with the general cluster metallicity [Fe/H]$\approx$-0.7 considering
an enhancement by 0.3 dex which is typical at these metallicities
(Edvardsson et al. 1993).
  
\begin{table}
\caption{Effective temperatures, measured equivalent widths
 and Li abundances. Errors in the EW  are 1$\sigma$. Errors 
in  Li abundances reflect only errors in EW. }
\parskip0.2cm
\begin{center}
\tabcolsep0.2cm
\begin{tabular}{r|r|r|r|r}\hline
\multicolumn{1}{c}{Star} &
\multicolumn{1}{c}{T$_{eff}$} &
\multicolumn{1}{c}{E.W.}&
\multicolumn{1}{c}{S/N}&
\multicolumn{1}{c}{N(Li)}
\\ \hline\hline
b7    & 5899   & 53  $\pm$ 8     & 40  & 2.41 $\pm$0.08  \\
b6    & 5823   & $<$40          & 23  & $<$ 2.19        \\
b5    & 5823   & 56  $\pm$ 11  & 30 & 2.37 $\pm$0.11  \\
b1    & 5030   & $<$10          & 30  & $< 0.73$          \\
L2734 & 3940   & $<$10          & 80  & $<$-0.70   \\
b7+b5 & 5866   & 52  $\pm$ 6.7   & 50 & 2.37 $\pm$ 0.07           \\
\hline
\end{tabular}
\end{center}
\end{table}

The T$_{eff}$ vs. M$_v$ diagram for the observed stars
is given in Figure2.

\section{Discussion}

\subsection{Comparison with  PopII stars and globular clusters}
\smallskip

In Table 3 temperatures, equivalent widths and Li abundances for the 
stars observed in 47Tuc are given. The errors in the Li abundances 
include only  uncertainties coming from the equivalent widths 
measurements but an  other comparable source of uncertainty
comes from  the error 
in the stellar T$_{eff}$ determination. Considering the  
uncertainty ($\pm$ 0.02) in the reddening
and possible systematic effects in the temperature scale adopted, 
an uncertainty  in  T$_{eff}$ of $\pm$ 100 K is assumed, which would correspond
to $\pm$0.075 dex in [Li]. Other sources of uncertainty in the Li determination
come from the errors in the adopted gravity and microturbulence, 
and their contribution is only of 0.02 dex, much lower than the equivalent 
width and effective temperature terms.  

Therefore the Li abundance on the plateau of 47Tuc
is:
\bigskip

\centerline{ [Li]= 2.37$\pm$ 0.15}

\bigskip
 The [Li] abundance in 47Tuc  is consistent with the {\it plateau} level 
of [Li]= 2.19 $\pm$ 0.016 
(Bonifacio and Molaro (1996)) at  
1.2 $\sigma$ level, but 
the Li level measured in 47Tuc is possibly slightly higher than the {\it plateau}.
This would imply  a moderate Li production for 47Tuc, which is not 
surprising considering the 
higher metallicity of 47Tuc.

Observations exist for the globular clusters NGC 6397 
(Molaro and Pasquini 1994; Pasquini and Molaro 1996) and M92 
(Deliyannis et al 1995), which have [Fe/H]=-2 and -2.25 respectively.
The mean value for the abundances
so far observed in stars close to the turnoff 
(ignoring upper limits) are [Li]=2.3 and 2.13 for NGC6397 
and M92, respectively.
These  values are, within the errors, consistent with the plateau
abundance and slightly lower than in 47Tuc.

In Figure 2 the stars observed in 47Tuc (triangles) are overimposed with
those observed in NGC 6397 (squares), after scaling the 
stellar apparent luminosities according to the respective cluster 
distances and reddening (Alcaino et al. 1987, Hesser et al. 1987),
and the Li abundances are given. 
In the same Figure the observations of M92 from Deliyannis et al. (1995)
(hexagons) are included. For M92 a distance module of 14.75 and a 
reddening of E(B-V)=0.03 have been assumed. 
The effective  temperatures of the stars in M92 and NGC6397 have been computed 
using the (Alonso et al. 1996) scale adopted for  47Tuc. 

The  Li abundances for M92 were recomputed from 
Deliyannis et al. (1995) Li equivalent widths using the same 
atmospheric models adopted in 
this work and  for NGC 6397. The  clusters are therefore in the same effective 
temperature reference scale and also the abundances are derived  
with the same atmospheric models.         
Measurements errors are affecting the
diagram of Figure 2 in a complex way. 
On the top of known uncertainties, those from  the 
(B-V) photometry will change slightly the stellar effective temperature,
not only giving different Li abundances, but also 
 moving the stars in the diagram.
 
The overall  similarity in the Li abundance among 47Tuc, the other globular 
clusters and the {\it plateau} is  remarkable, confirming that 
essentially the same Li abundance is observed over three  orders 
of magnitude variation in the metallicity. 
If, as suggested by Boesgaard (1991) primordial Li was an order 
of magnitude higher and the Li observed in these clusters has
been depleted  by diffusion, winds,
rotational mixing or other mechanism, then  our observations require 
these  mechanisms to be highly metallicity independent 
in the range -3.5 $\le$[Fe/H]$\le$-0.7. 
The same mechanism has instead to 
be  strongly  metallicity dependent at higher metallicity in order 
to reproduce the smooth increase of the Li abundance envelope 
observed in  the Li-Fe diagram of field stars  (Rebolo et al. 1988,
see also next section ).
On the other hand, if these stars  are essentially undepleted, the difference
can be easily accomodated by a modest Galactic Li production, with
most of the Galactic Li production occurring later on.

\subsection{Comparison with  PopI stars}
\smallskip

Several field stars  with measured Li abundance exist having 
metallicities similar to those of 47Tuc. 
In Figures 3a,b  observations collected from several authors
are  plotted. The  compilation 
includes all the field stars observed in Li with a  metallicity 
between -0.4 and -1,
 bracketing the metallicity of 47Tuc, with effective temperatures
higher than 5500 K and estimated gravities $\ge$3.5, to avoid object which 
may have undergone depletion or dilution.

It is known that at these metallicities a  real  spread in Li abundances 
exists among field stars  (Rebolo et al. 1988).
 The spread  is present at  temperatures  far from those 
of  the   `lithium dip' or  from those where  convective induced 
depletion occurs.
The spread is also unlikely  an evolutionary effect, because the stars are 
quite firmly classified as genuine dwarfs.  
This spread has been  interpreted in terms of different degrees 
of depletion and/or in terms of an age spread in the sample. 
One could expect   that older stars are on average the more 
depleted ones.

Thus, it is  remarkable that none of the field stars, 
which are on average younger than 47Tuc, 
have abundances significatively higher than those observed in 
47Tuc. Only HD 14221 (Balachandran  1990) with a metallicity 
of -0.4 and [Li]=2.9 is definitively higher, but, due to its high metallicity,
this object cannot be taken as a safe counterexample.  

The upper values of the observed [Li] both in field stars 
{\it and} 47Tuc at $\sim$ 2.4  indicate that the [Li] vs. Fe abundance 
curve given by the upper envelope of Figures 3a,b is basically 
independent of age, weakening the interpretation that an age-dependent 
depletion mechanism is in force (Boesgaard 1991).  

These upper values  are easily understood 
if no depletion has affected these stars.
This in turn would suggest that
we are just   observing the increase of the Galactic Li enrichment.   
It would imply that  only a very marginal increase in Li abundance has
occurred in the Galaxy from the primordial value up to metallicities 
of $\sim$ -0.5. In addition, since among field stars several are
expected to be younger than 47Tuc, this may also suggest that Li enrichment
depends more on metallicity rather  than on age. 
We have to remember, however, that  in old Pop I stars some 
age-independent mechanism is acting, which increases the scatter of
[Li] abundances with respect to the metal poor objects (Pasquini et al. 
1994), and this mechanism has not yet been satisfactorily identified.

According to the Be measurements in metal poor stars  
and the expected Li/Be yields, the observed enhancement can be explained by
high energy Cosmic Ray production, without the need of invoking 
other mechanisms. In fact, according to  
Molaro et al (1996), at [Fe/H]$\approx$-0.8, Be is  [Be]=0.94
and the corresponding inferred Li by the same spallation processes 
is [Li]=1.9, when only $^7$Li isotope is considered. This value 
added to a  primordial value of [Li]=2.20 gives  [Li]=2.38, which is
precisely what it is observed in 47Tuc.

Since an abundance up to [Li]$\sim$3.1 is observed in the 4 Gyrs old, 
solar metallicity cluster M67 (Pallavicini et al. 1996),
this interpretation would imply  an 
increase of Li abundance by 0.7 dex in the time (and metallicity) 
 interval between 47Tuc and M67. 
 
\begin{figure}
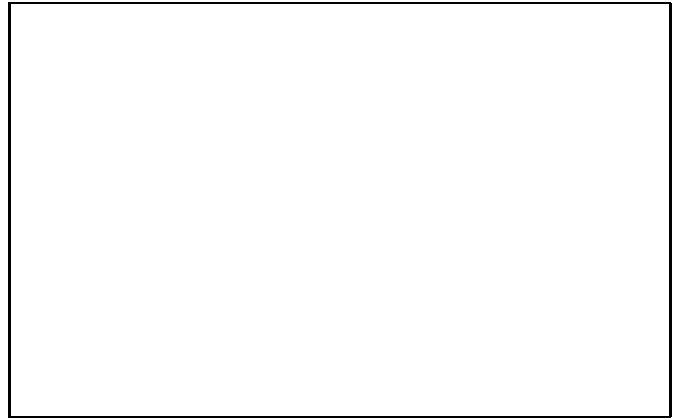

\picplace{5.5cm}
\caption[2]{Absolute magnitudes and  Li abundances vs. 
T$_{eff}$ for the 47Tuc stars (Triangles), the stars observed 
in NGC6397 (squares) and M92(exagons). 
The apparent stellar magnitudes
were corrected for the distance modulus of the cluster and for reddening.
The derived Li abundances are given.}
\end{figure}

\begin{figure*}
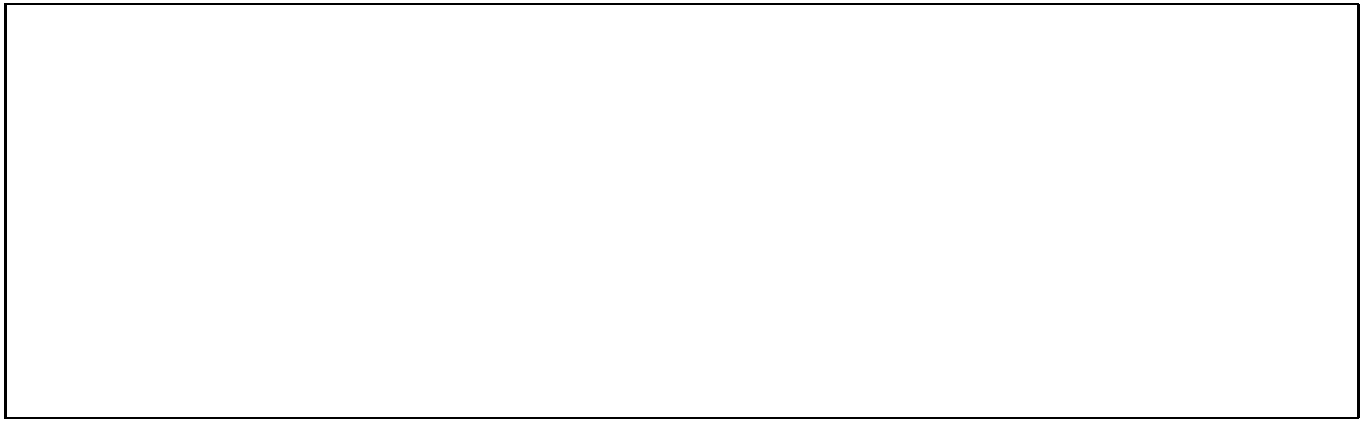

\picplace{5.5cm}
\caption[3]{3a): Li abundance vs. T$_{eff}$ for the turnoff stars of 
47Tuc  and field dwarfs having similar metallicities to the cluster.
3b) Li abundance vs. metallicity, for the same sample of 3a).}
\end{figure*}

\subsection{Li dispersion in 47Tuc}
\smallskip
 
For the third turnoff star (b6) only an upper limit  could be obtained.
If the presence of Li abundance scatter among 47Tuc stars
will be confirmed, this would indicate that the scatter observed in 
field stars at the 47Tuc metallicities is not due to age or chemical 
composition differences, but rather to  an extra  mechanism capable of 
overdepleting Li.

It should be noticed that single stars in each cluster  present 
Li abundances which are as high as $\sim$2.4. In presence
of a real dispersion of the Li abundances in the clusters stars, 
only these high values would be  close to the initial one.
 
The presence of a spread in Li abundances for similar stars 
belonging to M92  has been claimed by Deliyannis et al. (1995)
and successive observations (Boesgaard et al. 1996) confirmed 
that a statistically significative difference in the measured Li 
equivalent widths exists. 
The three M92 stars considered
are subgiants, and they have an effective temperature around 
5750 K, that is very close to the point where subgiant 
dilution begins to be
present (Pilachowski et al. 1993). These stars   are just 100 K hotter
than star C4044 in NGC6397, where the signs of  dilution are evident 
(cfr. figure 2). Photometric uncertainties may play an important role.
An uncertainty in the (B-V) colours of 0.02, as given by 
Stetson and Harris (1988) for two of the stars observed,
corresponds to a difference in temperature of almost 100 K. 
This error  is  not only relevant for the derived abundances ($\pm$ 0.07),
but it may also shift 
the stars into the temperature regime   where dilution is observed. 
A lower reddening, still allowed by the photometric observations,
would result into lower effective temperatures, also moving the stars
closer to the dilution zone.
In addition, Boesgaard  et al. (1996) observed for the 
Li-rich subgiant also different Mg lines, and this could indicate that 
other intrinsic differences exist among the M92 stars. 

In 47Tuc the comparison between star b6  with   stars b5 and b7  may also
suggest the presence of a scatter.
For this case the stars have a temperature too hot to 
expect that depletion or dilution are effective, 
but with the S/N ratio of our observations
the possible difference cannot be considered as definitive. 
It is  worth noticing that 
stars b5 and b7 have the same Li and they are respectively
CN rich and CN poor according to 
Briley et al. (1994); therefore CN inhomogeneities among main sequence stars
of 47Tuc may not be related to the dispersion of Li. The independence
of the Li abundance of nitrogen abundance had already been found by Spite and 
Spite (1986) by  studying a sample of nitrogen rich field stars, and our result
confirms Spite's conclusion that  nitrogen enhancement  
is not produced by deep mixing.

Pending more accurate observations,  we can 
only conclude at the moment that the spread 
among similar stars of the same cluster has to be confirmed.

\subsection{Evolved stars}

Very low Li abundances are derived for the evolved stars in 47Tuc
and NGC6397.
 This conclusion is  not affected 
by the considered uncertainties. As pointed out by Pilachowski et al. (1993)
and Pasquini and Molaro (1996), all present theories predict a 
levelling of the Li abundance at [Li]$\sim$1.1 
for evolved stars with temperature
below  $\sim$ 5100 K (Swenson 1995), while observations in field 
and NGC6397 subgiants show a dramatic drop. 
Our tight upper limits  on b1 and L2734
in 47Tuc confirm these previous studies, indicating that 
some other mechanism should be present, in addition to dilution, 
to produce an extra Li depletion when stars are ascending the RGB.

\section{Conclusions}

In this work we have presented and analyzed the first Li observations in 5
turnoff and evolved stars of the old, metal rich globular cluster 47Tuc. 
Pushing the NTT telescope to its limits we have obtained 
high resolution observations for three  V$\sim$ 17.4 stars located 
in the Spite {\it plateau}.

The main results are:

1) The Li abundance derived in two of the turnoff stars and from their 
composite spectrum is 2.37 $\pm$ 0.15. This value is slightly higher
than the 2.20 found in the Pop II stars suggesting at most a  mild increase
in the Li abundance. Spallated Li as deduced from the observed [Be]
can account for such an increase.

 2) This finding, together with the fact that no field star in the metallicity 
range -1 to -0.4 show Li levels substantially 
higher than the 47Tuc turnoff stars,
 pose strong requirements  on the possible models of 
Li depletion in the stellar interior. Proposed models have to 
account for an almost constant Li level which  extends from the most 
metal poor stars known up to metallicity  -0.7.

3) For the third turnoff star only an upper limit  could be obtained.
If the presence of Li abundance scatter among 47Tuc stars were
 confirmed, this would indicate that the scatter observed in 
field stars with similar  metallicities is not due to age or chemical 
composition differences, but rather to  an extra  mechanism capable of 
overdepleting Li.

4) The tight upper limits for the 2 evolved stars confirm the finding that
some extra depletion has to be present in evolved stars in addition to 
dilution. 

\acknowledgements{  We thank F. Spite for providing us with a database 
of Li observations of field stars, and for his peer review.
We thank Piercarlo Bonifacio for his advice
in the Li abundance determination}


\begin{thebibliography}{}

\bibitem[1987]{al}
Alcaino, G., Buonanno, R., Caloi, V., Castellani, V., Corsi, C.E., 
Iannicola, G., Liller, W. 1987 AJ 94, 917.
\bibitem[1996]{ala}
Alonso, A., Arribas, S., Martinez-Roger, C. 1996 A\&AS 117, 227
\bibitem[1996]{alb}
Alonso, A., Arribas, S., Martinez-Roger, C. 1996 A\&A 313, 873
\bibitem[1990]{bal}
Balachandran, S. 1990 ApJ 354, 310.
\bibitem[1993]{ba}
Balachandran, S. 1995 ApJ 446, 203.
\bibitem[1988]{bn}
Banse, K., Grosbol, P., Ponz, D. et al 1988 `The MIDAS Image Processing System'
in ` Instrumentation for Ground Base Astronomy: present and Future', 
Robinson, L.B. ed. (Springer)
\bibitem[1996]{boes2}
Boesgaard, A.M. 1996 in `Formation of the Galactic Halo....Inside and 
Out', H. Morrison and A. Sarajedini eds. p. 327. ASP Conf. Ser. 92
\bibitem[1991]{boes1}
Boesgaard, A.M. 1991 ApJ 370, L95. 
\bibitem[1996]{bm}
Bonifacio, P. Molaro, P. 1996 MNRAS in press.
\bibitem[1994]{br1}
Briley, M.M., Hesser, J.E., Bell, R.A. 1991 ApJ 373, 482. 
\bibitem[1994]{br2}
Briley, M.M., Hesser, J.E., Bell, R.A., Bolte, M., Smith, G.H. 1994 AJ 108, 
2183.
\bibitem[1989]{bro}
Brown, J.A., Sneden, C., Lambert, D.L., Dutchover, E. Jr. 1989 ApJS 71, 293.
\bibitem[1992]{bc}
Buser, R., Kurucz, R.L. 1992 A\&A 264, 557.
\bibitem[1988]{cay}
Cayrel, R. 1988, in `IAU Symp. 132, The impact of very high S/N spectroscopy 
on stellar physics', G. Cayrel de Strobel and M. Spite eds. p. 345
(Kluwer)
\bibitem[1974]{cann}
Cannon R. D. 1974 MNRAS 167 551.
\bibitem[1995]{cra}
Crane, P. 1995 `The Light Element Abundance' (Springer).
\bibitem[1975]{crow}
Crawford, D.L., Snowden, M.S. 1975 PASP 87, 561.
\bibitem[1991]{dm}
D'Antona, F., and Mazzitelli, I. 1994 ApJS 90, 467.
\bibitem[1995]{dbk}
Deliyannis, C.P., Boesgaard, A.M., King, J.R. 1995 ApJ 452, L13. 
\bibitem[1985]{dod}
D' Odorico, S., Gratton, R.G., Ponz, D. 1985 A\&A 142, 232.
\bibitem[1993]{edv}
Edvardsson B., Andersen, J., Gustafsson, B., Lambert, D.L., Nissen, P.E., 
Tomkin, J. 1993 A\&A 275, 101
\bibitem[1992]{gi}
Giraud, E. 1995 `The ESO Multi-Mode Instrument
        and The Superb Seeing Imager', ESO Operating Manual.
\bibitem[1987]{hes}
Hesser, J.E., Harris, W.E. et al. 1987 PASP 99, 739.  
\bibitem[1976]{hesp}
Hesser, J.E., Philip A. G. D. 1976 PASP 88 89.  
\bibitem[1988]{hp}
Hobbs, L.M., Pilachowski 1988 ApJ 334, 745.
\bibitem[1993]{ku}
Kurucz, R. L. 1993 CD-ROM No 11, 13, 18.
\bibitem[1977]{l}
Lee S. W. 1977 A\&AS 27, 381
\bibitem[1991]{mm}
Mayor, M., Imbert, M. et al. 1984 A\&A 134, 118.
\bibitem[1973]{men}
Menzies J. 1973 MNRAS 163, 323
\bibitem[1991]{mc}
Michaud G., Charbonneau P. 1991 Space Sci. Rev 57,1.
\bibitem[1994]{mp}
Molaro, P., Pasquini, L.  1994 A\&A 281, L77.
\bibitem[1995]{mo}
Molaro, P., F. Primas, Bonifacio, P.  1995 A\&A 295, L47.
\bibitem[1996]{ppr}
Pallavicini, R., Pasquini, L., Randich, S. 1996 A\&A in preparation.
\bibitem[1994]{plp}
Pasquini, L., Liu, Q., Pallavicini, R. 1994 A\&A 287, 191.
\bibitem[1995]{pm}
Pasquini, L., Molaro, P. 1996 A\&A 307, 761. 
\bibitem[1992]{pi}
Pinsonneault, M.H., Deliyannis, C. P., \& Demarque, P. 1992 ApJS 78, 179.
\bibitem[1993]{pila}
Pilachowski, C.A., Sneden, C., Booth, J. 1993 ApJ 407, 699. 
\bibitem[1996]{rgpp}
Randich, S., Gratton, R., Pallavicini, R., Pasquini, L. 1996 
A\&A in preparation.
\bibitem[1988]{rmb}
Rebolo. R., Molaro P., Beckman, J. E., 1988 A\&A 192, 192
\bibitem[1993]{so}
Soderblom, D. R. et al 1993 AJ 106, 1059
\bibitem[1986]{sn}
Spite, F., Spite, M. 1986, A\&A 163, 140
\bibitem[1995]{spf}
Spite, F. 1995 in `The Light Element Abundance', P. Crane ed., 239 (Springer)
\bibitem[1995]{swe}
Swenson, F.J. 1995 ApJ 438, L87
\bibitem[1991]{tri}
Trimble, V. 1991 A\&AR 3, 1
\bibitem[1985]{vdbb}
Vandenberg, D.A.,  Bell, R.A. 1985 ApJS 58, 561
\bibitem[1995]{vc}
Vauclair, S., Charbonnel, C. 1995 A\&A 295, 715
\bibitem[1994]{zah}
Zahn, J.P. 1994 A\&A 288, 829.
\end{thebibliography}
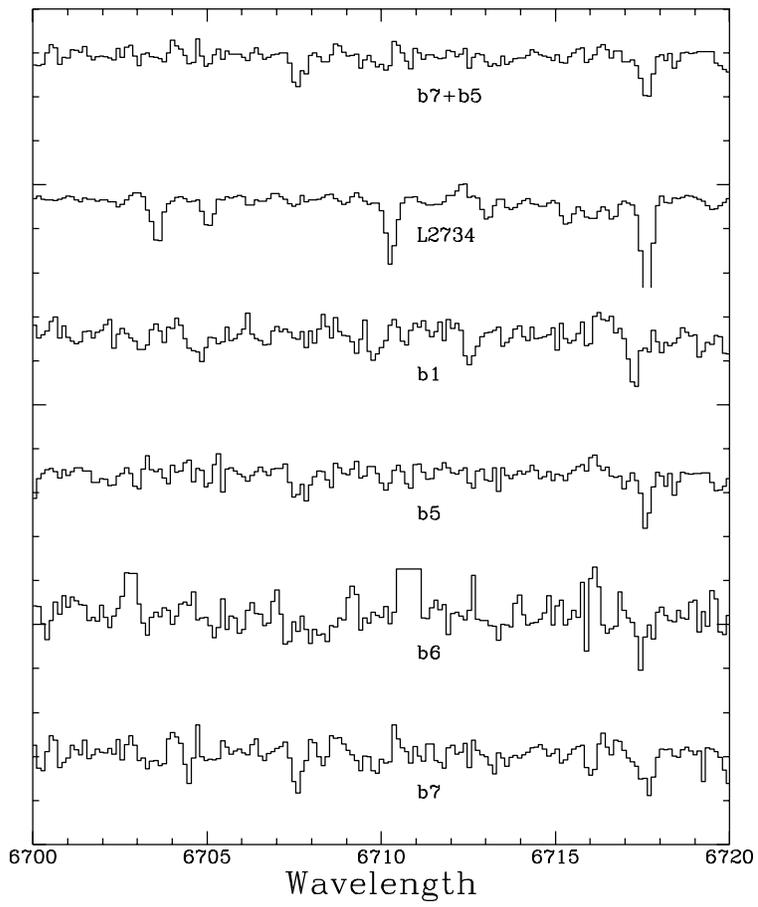
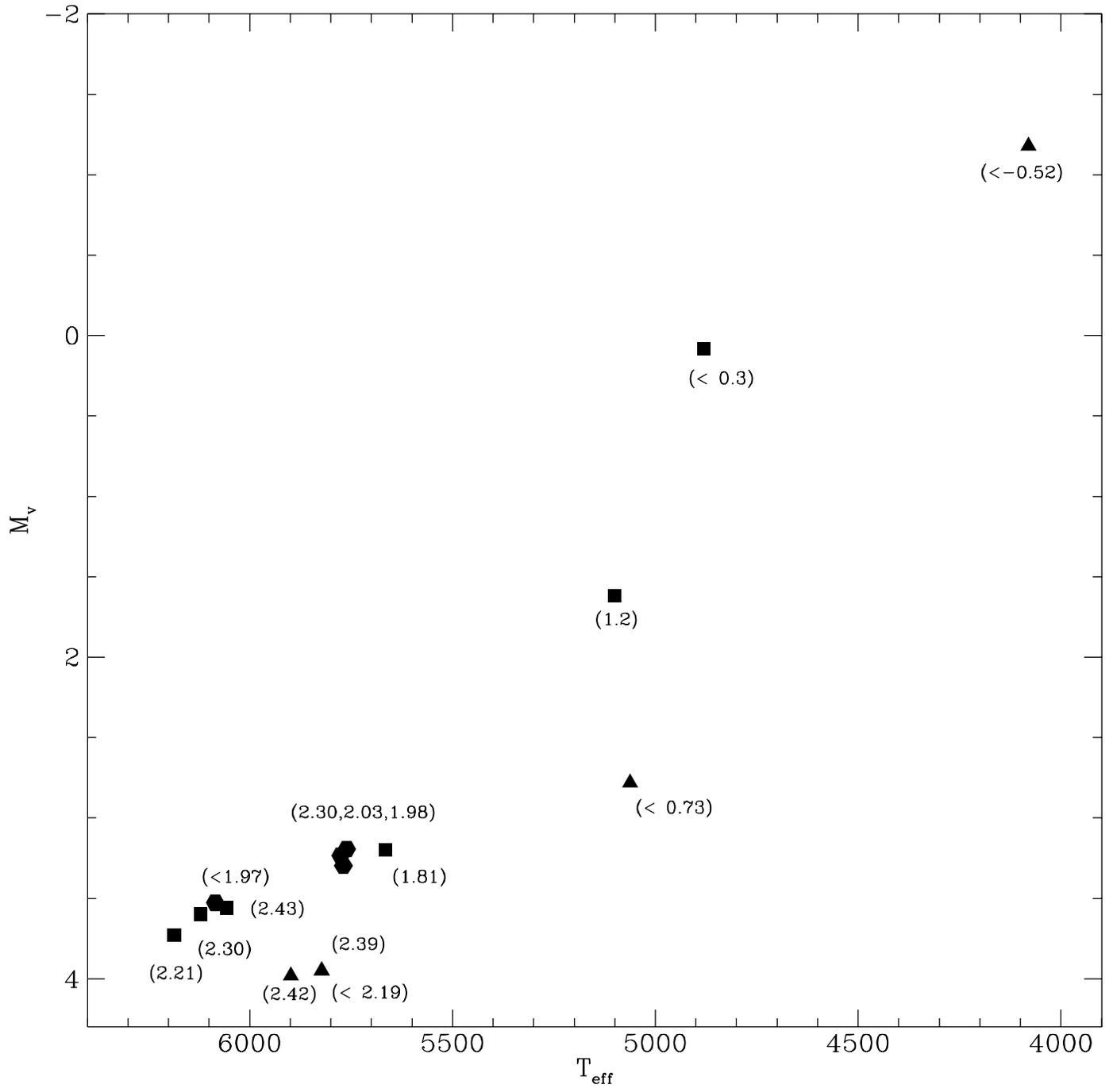
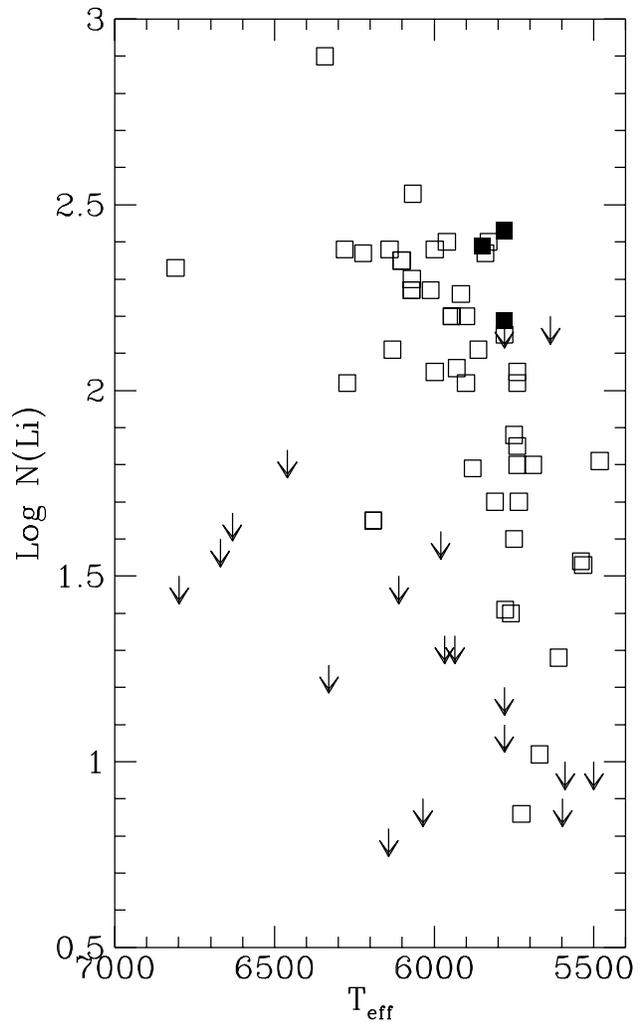
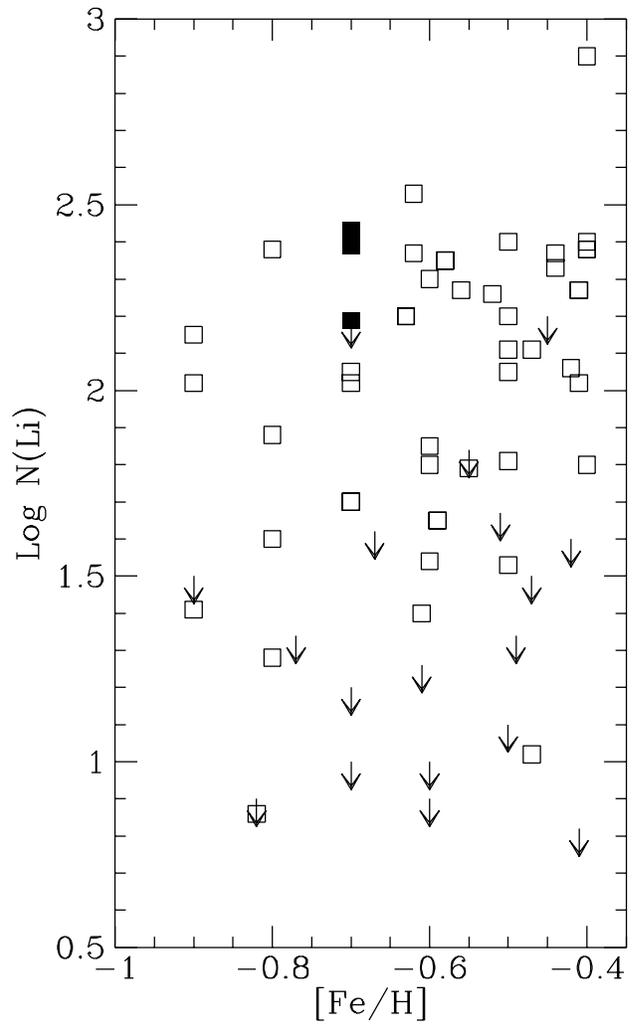
\end{document}